\documentclass[a4paper,11pt]{article}
\usepackage{pos}

\usepackage{epsfig}
\usepackage{multicol}
\usepackage{pstricks,pst-grad}
\usepackage{feynmf}

\newcommand{\no}{\noindent}
\newcommand{\myeq}[3]{\vspace{#2} \begin{equation} \hspace{#1} #3 \end{equation} \vspace{0cm}}

\newcommand{\tx}[1]{\mathrm{#1}}

\newcommand{\La}{\mathcal{L}}
\newcommand{\Mq}{\mathcal{M}}
\newcommand{\Op}{\mathcal{O}}
\renewcommand{\bf}[1]{\textbf{\boldmath #1}}

\title{Three-flavour order parameters of chiral symmetry in low-energy QCD}

\author*[a]{Mari\'{a}n Koles\'{a}r}
\author[a]{Ji\v{r}\'{i} Novotn\'{y}}

\affiliation[a]{Institute of Particle and Nuclear Physics, Faculty of Mathematics and Physics, Charles University,\\ V Holešovičkách 2, Prague, Czech republic}

\emailAdd{kolesar@ipnp.troja.mff.cuni.cz}
\emailAdd{novotny@ipnp.troja.mff.cuni.cz}

\abstract{The current state of knowledge of the order parameters of the spontaneously broken chiral symmetry, the quark condensate and the pseudoscalar decay constant in the chiral limit, is briefly reviewed, based on available phenomenological fits and lattice QCD calculations. We argue that while the theory is pretty well understood in the two-flavour case, there is still a gap in the knowledge of the characteristics of the QCD vacuum in the three-flavour one. Our constraints for the three-flavour parameters obtained by a Bayesian statistical analysis of the decays of eta to three pions are presented. Possible implications of a new analysis of subthreshold parameters of pion-pion scattering is outlined.}

\FullConference{%
  40th International Conference on High Energy physics - ICHEP2020\\
  July 28 - August 6, 2020\\
  Prague, Czech Republic (virtual meeting)
}

\begin{document}
\maketitle

\section{Introduction}

Quantum chromodynamics (QCD), as a theory of strong interactions, is constructed using the principle of local gauge invariance. The fundamental degrees of freedom of the theory are the quark and gluon fields, the gauge symmetry is the colour $SU(3)$ group. As is well known, the coupling constant $g$ grows at low energies due to gluon self-interactions and the usual perturbation expansion of quantum field theory fails. The quarks and gluons are thus not the appropriate degrees of freedom at low energies, the quarks are confined and the hadron spectrum is born.

One of the approaches used to tackle this phenomenon is the framework of effective field theory. In this framework, an effective Lagrangian is constructed as the most general form respecting the symmetries of the original theory, with the appropriate low-energy degrees of freedom being used. 

Such a program for describing the physics of strong interactions at low energies was proposed by Weinberg \cite{Weinberg:1978kz}, now called Chiral perturbation theory ($\chi$PT). The governing symmetry group is the approximate chiral symmetry $SU(N_f)_L\times SU(N_f)_R$ of the QCD Lagrangian, where $N_f$ is the number of quark flavours considered light. This symmetry is spontaneously broken and the relevant degrees of freedom at low energies are pseudoscalar mesons - the Goldstone bosons of the spontaneously broken chiral symmetry. These can be identified with the pions in the two flavour case or the octet of pions, kaons and eta in the three flavour one. Their fields can be collected in a matrix field $U(x)$, the basic building blocks of the theory then are: 

\myeq{0cm}{0cm}{
		U(x)\, =\, e^{\frac{i}{F(N_f)}\,\phi^a(x)\lambda^a},\quad 
		\Mq\, =\, \tx{diag}(m_u,m_d[,m_s]),}
									
\no where $\phi^a$ is the triplet/octet of the pseudoscalar mesons, $\lambda^a$ the Pauli/Gell-Mann matrices and $m_q$ the quark masses. $U(x)$ transforms conveniently under the chiral symmetry \cite{Coleman:1969}

\myeq{0cm}{0cm}{
		U'(x) \ =\ U_R U(x) U_L^+ \,,\quad \Mq'\ =\ \ U_R \Mq U_L^+, \quad 
		U_{L,R}\in SU(N_f)_{L,R}.}

The effective Lagrangian is organized as an expansion in derivatives and quark masses, using the power counting \cite{Weinberg:1978kz}

\myeq{0cm}{0cm}{D = 2 + 2L + \sum_n V_n(n-2),}

\no with $D$ being the chiral dimension of a diagram with $L$ loops and $V_n$ vertices of order $\Op(p^n)$. In addition, the quark masses are counted as $m_q$$\,\sim\!O(p^2)$ in the standard version of the power counting.	The theory then contains an infinite number of effective coupling constants, called low energy constants (LECs), but is renormalizable order by order. The range of validity of the effective theory is bounded from above by the mass of the lightest state not explicitly included, which is the mass of kaon in the two-flavour case and the mass of the $\rho$ meson in the three-flavour one.

In this picture, the theory was worked out up to next-to-next-to-leading order \cite{Gasser:1983yg,Gasser:1984gg,Bijnens:1999sh,Bijnens:2001bb}. The framework was also extended further in several directions, but that is beyond the scope of our brief review. At the leading order, there are two independent contributions 

\myeq{0cm}{0cm}{
		\La^{(2)}\ =\ \frac{F(N_f)^2}{4} \tx{Tr}[D_{\mu}U D^{\mu}U^+ +\, 2B(N_f)(U^+ \Mq + \Mq^+ U)].}
		
At the next-to-leading order, the (phenomenologically relevant) two-flavour LECs are $l_1-l_7$, the three-flavour ones are $L_1-L_{10}$. At the next-to-next-to-leading order, the two-flavour constants are $c_1-c_{52}$ and $c_1^W-c_{13}^W$, while the three-flavour ones are $C_1-C_{90}$ and $C_1^W-C_{23}^W$.

The leading order constants $F$$\equiv$$F(2)$, $B$$\equiv$$B(2)$ and $F_0$$\equiv$$F(3)$, $B_0$$\equiv$$B(3)$	are sometimes referred to as order parameters of the spontaneously broken chiral symmetry. $F(N_f)$ is the pseudoscalar decay constant in the chiral limit, $B(N_f)$ is related to the quark condensate in the chiral limit $\Sigma(N_f)=B(N_f) F(N_f)^2$. We will use a convenient normalization for the order parameters

\myeq{0cm}{0cm}{Z(N_f) = \frac{F(N_f)^2}{F_{\pi}^2},\quad
								X(N_f) = \frac{2\,\hat{m}\Sigma(N_f)}{M_{\pi}^2 F_{\pi}^2} 
												   = \frac{2\,\hat{m}B(N_f) F(N_f)^2}{M_{\pi}^2 F_{\pi}^2},}
													
\no where $M_\pi$ is the pion mass, $F_\pi$ is the pion decay constant and $\hat{m}$$=$$(m_u+m_d)/2$. In this normalization, the value of $Z(N_f)$ and $X(N_f)$ is in the range $(0,1)$. A value close to one signifies the dominance of the leading order in the chiral expansion and thus a good convergence of the theory. A value significantly smaller than one could be problematic for the theory, as irregularities in the expansion might arise due to the leading order not being sufficiently dominant.

In section 2 we briefly review the current state of knowledge of the leading order LECs from two sources - phenomenological fits and numerical solutions of QCD on a lattice (lattice QCD). We highlight the unsatisfactory situation in the three flavour case. In section 3 we present our attempt to constrain the three-flavour order parameters from $\eta\to 3\pi$ decays and discuss the connection with the result from $\pi\pi$ scattering. We conclude in section 4.

\section{State of knowledge of the order parameters}

Table \ref{tab_2-fl} contains a few selected results for the two-flavour order parameters. The Flavour Lattice Averaging Group (FLAG) provides an immensely helpful service of assessing the output of various lattice QCD collaborations and we rely on their most recent review \cite{Aoki:2019cca}.

As can be clearly seen, the values of the leading order LECs are pretty close to one and thus dominate the chiral expansion. 

\begin{table}[h] \small \begin{center}
\begin{tabular}{|c|c|@{\,}|c|c|}
  \hline \rule[-0.2cm]{0cm}{0.5cm} phenomenology & source & $Z(2)$ & $X(2)$ \\
	\hline \rule[-0.2cm]{0cm}{0.5cm} Descotes et al.'01 \cite{DescotesGenon:2001tn} & $\pi\pi$ scattering & 0.89$\pm$0.03 & 0.81$\pm$0.07 \\
	\hline \hline \rule[-0.2cm]{0cm}{0.5cm} lattice QCD & source & $Z(2)$ & $X(2)$ \\
	\hline \rule[-0.2cm]{0cm}{0.5cm} Bernard et al.'12 \cite{Bernard:2012fw} & RBC/UKQCD & 0.86$\pm$0.01 & 0.89$\pm$0.01 \\
	\hline \rule[-0.2cm]{0cm}{0.5cm} FLAG'19 $N$=2+1 \cite{Aoki:2019cca} & average & 0.89$\pm$0.01 & 0.84$\pm$0.05 \\
	\hline \rule[-0.2cm]{0cm}{0.5cm} FLAG'19 $N$=2+1+1 \cite{Aoki:2019cca} & average & 0.862$\pm$0.005 & 0.98$\pm$0.24 \\
	\hline
\end{tabular}\end{center}
	\caption{Selected results for the two-flavour order parameters.}
	\label{tab_2-fl}
\end{table}\normalsize

Table \ref{tab_3-fl} contains selected results for the three-flavour order parameters. As can be noticed immediately, the values vary widely. FLAG provides no average, as only one collaboration (MILC) has a determination of high standing according to their assessment. It should be noted that this determination is now more than a decade old. We have also listed results which employ very similar lattice QCD data within distinct theoretical frameworks (resummed $\chi$PT and large $N_c$). And it can be observed that different theoretical assumptions lead to some very different outcomes. 

\begin{table}[h] \small \begin{center}
\begin{tabular}{|c|c|@{\,}|c|c|}
	\hline \rule[-0.2cm]{0cm}{0.5cm} phenomenology & source & $Z(3)$ & $X(3)$ \\
	\hline \rule[-0.2cm]{0cm}{0.5cm} Bijnens, Ecker'14 \cite{Bijnens:2014lea} & NNLO $\chi$PT (``main fit'') & 0.59 & 0.63\\
	\hline \rule[-0.2cm]{0cm}{0.5cm} Bijnens, Ecker'14 \cite{Bijnens:2014lea} & NNLO $\chi$PT (``free fit'') & 0.48 & 0.45 \\
	\hline \rule[-0.2cm]{0cm}{0.5cm} Amoros et al.'01 \cite{Amoros:2001cp} & NNLO $\chi$PT (``fit 10") & 0.89 & 0.66 \\
	\hline\hline \rule[-0.2cm]{0cm}{0.7cm} lattice QCD & source & $Z(3)$ & $X(3)$ \\
	\hline \rule[-0.2cm]{0cm}{0.5cm} Bernard et al.'12 \cite{Bernard:2012ci} & RBC/UKQCD+Re$\chi$PT & 0.54$\pm$0.06 & 0.38$\pm$0.05\\
	\hline \rule[-0.2cm]{0cm}{0.5cm} Ecker et al.'13 \cite{Ecker:2013pba} & RBC/UKQCD+large $N_c$ & 0.91$\pm$0.08 & \\
	\hline \rule[-0.2cm]{0cm}{0.5cm} FLAG'19 $N$=2+1 \cite{Bazavov:2009fk} & MILC 09A & 0.72$\pm$0.06 & 0.61$\pm$0.06 \\
	\hline
\end{tabular}\end{center}
	\caption{Selected results for the three-flavour order parameters.}
	\label{tab_3-fl}
\end{table}\normalsize

Pure phenomenological analyses, based on NNLO $\chi$PT fits, do not provide error bars and the values of various fits differ quite substantially. Overall, the situation seem quite unsatisfactory and we can only conclude that the values of the three-flavour order parameters are not known very well and values much smaller than one can not be excluded.

\section{Results: $\eta\to3\pi$ and $\pi\pi$ scattering} 

We have attempted to extract constraints on the three-flavour order parameters from experimental data on the $\eta\to 3\pi$ decays, details can be found in \cite{Kolesar:2017xrl}. We employed a Bayesian statistical approach within resummed $\chi$PT, an approach first used in \cite{DescotesGenon:2003cg} in the case of $\pi\pi$ scattering.

\begin{figure}
  \begin{center}
    \epsfig{figure=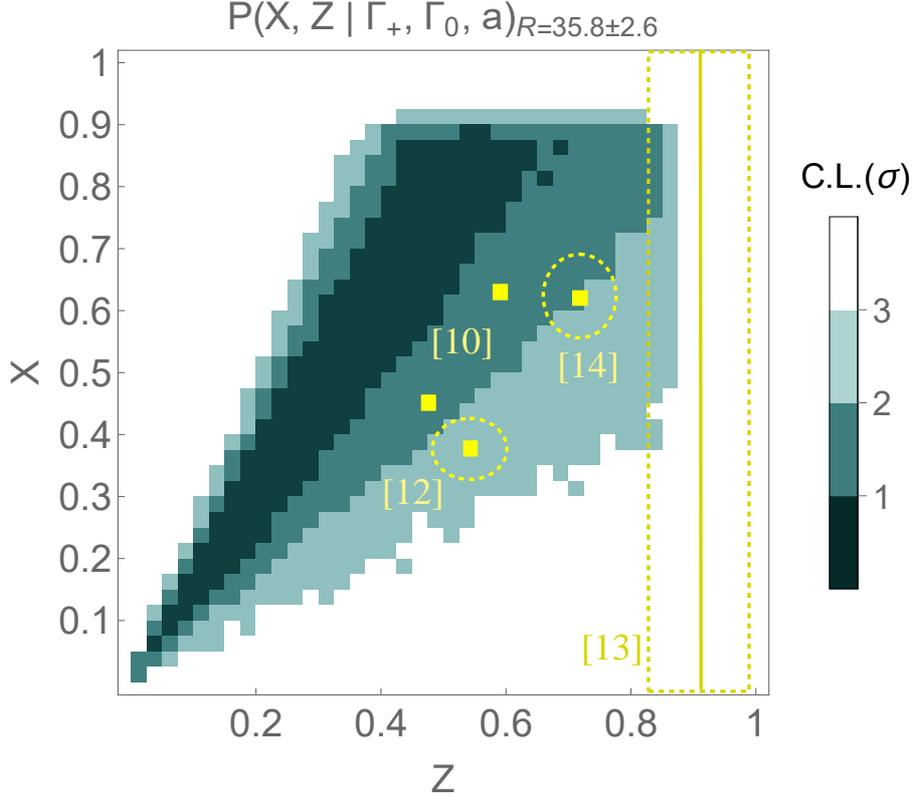,width=0.8\textwidth}
  \end{center}
	\caption{Probability density of the three-flavour order parameters from $\eta\to3\pi$ data (decay widths, Dalitz parameter $a$). Overlaid - results quoted in table \ref{tab_3-fl}.}
  \label{fig_X-Z}
\end{figure}

Our main result is depicted in figure \ref{fig_X-Z}, along with some of the determinations from table \ref{tab_3-fl}. Although we can exclude much of the parameter space at 2$\sigma$ confidence level, we are not able to achieve the same with any of the determinations we have listed, though some tension can be observed. We have obtained an upper bound for the chiral decay constant and a value for the ratio of the order parameters:

\myeq{0cm}{0cm}{Z<0.78\ (2\sigma\ \tx{CL}),\quad Y=\frac{X}{Z}=1.44\pm0.32.}

We have also reproduced the results for $\pi\pi$ scattering \cite{DescotesGenon:2003cg}. However, we have observed some tension between the $\eta\to3\pi$ and the $\pi\pi$ scattering data, which limited our ability to draw more definite conclusions. The possible source is the experimental error of the observables we used, especially the subthreshold parameter $\alpha_{\pi\pi}$. More recent data on the $\pi\pi$ scattering lengths, which can be used as an input, is available from the NA48/2 collaboration \cite{NA48:2010zza}, so we plan to update our analysis.

\section{Conclusion}

It was already understood a long time ago \cite{Appelquist:1998rb}, that varying the number of light quark flavours has a non-trivial effect on the phase structure of $SU(N)$ gauge theories, like QCD. At a low number of light flavours $N_f$, there is quark confinement and chiral symmetry breaking. However, at some critical number of flavours, $N_f$=$N_f^c$, a chiral phase transition occurs and chiral symmetry is restored. There is a conformal windows above that and at sufficiently high number of flavours, $N_f$$\geq$$N_f^A=11/2N_c$, asymptotic freedom is lost.

Of course, the critical number of light flavours has to be much larger than $N_f$=3. Still, a gradual suppression of the order parameters, the so-called paramagnetic inequality \cite{DescotesGenon:1999uh}, is expected to occur

\myeq{0cm}{0cm}{F(N_f+1)\, <\, F(N_f),\quad \Sigma(N_f+1)\, <\, \Sigma(N_f).}

\no Hence, there could be some difference between the behaviour of the $SU(2)$ and $SU(3)$ theories, possibly tied to large $N_c$ and Zweig rule violation in the scalar sector, as was already speculated two decades ago \cite{DescotesGenon:1999uh}. 

As we have discussed in this brief review, two decades later, the issue is still not solved. To our knowledge, the values of the three-flavour order parameters are not known to satisfactory precision and small values can not be safely excluded. \\	

\no \bf{Acknowledgment:} This work was financially supported by
The Czech Science Foundation (project GACR no.18-17224S).

\bibliographystyle{utphys}
\bibliography{Bibliography}

\end{document}